\documentclass[nohyper,11pt,letterpaper]{JHEP3}
\usepackage[dvips]{epsfig}
\usepackage{epsf,amsfonts,amssymb}

\newcommand{\be}{\begin{equation}}
\newcommand{\ee}{\end{equation}}
\newcommand{\ben}{\begin{displaymath}}
\newcommand{\een}{\end{displaymath}}
\newcommand{\bea}{\begin{eqnarray}}
\newcommand{\eea}{\end{eqnarray}}
\newcommand{\bean}{\begin{eqnarray*}}
\newcommand{\eean}{\end{eqnarray*}}

\newcommand{\ba}{\begin{array}}
\newcommand{\ea}{\end{array}}
\newcommand{\bi}{\begin{itemize}}
\newcommand{\ei}{\end{itemize}}

\def\g {\gamma}

\renewcommand{\L}{\Lambda}

\newcommand{\eg}{{\it e.g.,}\ }
\newcommand{\ie}{{\it i.e.,}\ }

\newcommand{\beq}{\begin{equation}}
\newcommand{\eeq}{\end{equation}}
\newcommand{\beqr}{\begin{displaymath}}
\newcommand{\eeqr}{\end{displaymath}}
\newcommand{\beqa}{\begin{eqnarray}}
\newcommand{\eeqa}{\end{eqnarray}}
\newcommand{\beqar}{\begin{eqnarray*}}
\newcommand{\eeqar}{\end{eqnarray*}}

\newcommand{\cK}{{\cal K}}

\newcommand{\labell}[1]{\label{#1}} 
\newcommand{\reef}[1]{(\ref{#1})}

\title{\LARGE Bouncing Braneworlds Go Crunch!}

\author{Jordan L. Hovdebo and Robert C. Myers\\
   Perimeter Institute for Theoretical Physics \\
         Waterloo, Ontario N2J 2W9, Canada \\
   Department of Physics, University of Waterloo \\
      Waterloo, Ontario N2L 3G1, Canada

E-mail: \email{jlhovdeb@sciborg.uwaterloo.ca,
rmyers@perimeterinstitute.ca}}

\abstract{Recently, interesting braneworld cosmologies in the
Randall-Sundrum scenario have been constructed using a bulk
spacetime which corresponds to a charged AdS black hole. In
particular, these solutions appear to `bounce', making a smooth
transition from a contracting to an expanding phase. By
considering the spacetime geometry more carefully, we demonstrate
that generically in these solutions the brane will encounter a
singularity in the transition region.}

\keywords{Braneworld Cosmology}

\preprint{\tt{hep-th/0308088}}

\begin{document}

\section{Introduction}

Braneworlds \cite{NAH,RS1,RS2} have generated enormous interest in
higher-dimensional spacetimes amongst particle theorists. A key
ingredient in these brane models is that the Standard Model
particles remain confined to a (3+1)-dimensonal brane, while only
the gravitational excitations propagate through the full
spacetime. Such scenarios provide a new framework in which to
address many longstanding puzzles in particle physics, such as the
hierarchy problem. The cosmology community has also shown an
increasing interest in braneworlds \cite{CS,elias}, since this is
another field where brane models have the potential to provide
novel solutions to many of the perennial questions.

In the present paper, we will focus on one small aspect of the
braneworld description of cosmology. In particular, we are
interested in a certain family of cosmological solutions
\cite{PEL} which were recently proposed in the context of the
Randall-Sundrum (RS) scenario \cite{RS1,RS2}. Recall that the RS
model introduces a codimension-one brane into a five-dimensional
bulk spacetime with a negative cosmological constant. The
gravitational back-reaction due to the brane results in
gravitational warping which produces massless graviton excitations
localized near the brane. Fine tuning of the brane tension allows
the effective four-dimensional cosmological constant to be zero
(or nearly zero). Brane cosmologies where the evolution is
essentially that of a four-dimensional Friedmann-Robertson-Walker
(FRW) universe can be constructed with a brane embedded in either
AdS \cite{CS} or an AdS black hole \cite{elias,KR}.

In either of the above cases, however, the cosmological evolution
on the brane is modified at small scales. In particular, if the
bulk space is taken to be an AdS black hole {\it with charge}, the
universe can `bounce' \cite{PEL}. That is, the brane makes a smooth
transition from a contracting phase to an expanding phase. From a
four-dimensional point of view, singularity theorems \cite{hawk}
suggest that such a bounce cannot occur as long as certain energy
conditions apply. Hence, a key ingredient in producing the bounce
is the fact that the bulk geometry may contribute a negative
energy density to the effective stress-energy on the brane
\cite{negative}. At first sight these bouncing braneworlds are
quite remarkable, since they provide a context in which the
evolution evades any cosmological singularities yet the dynamics
are still controlled by a simple (orthodox) effective action. In
particular, it seems that one can perform reliable calculations
without deliberating on the effects of quantum gravity or the
details of the ultimate underlying theory. Hence, several authors
\cite{others,higherd,kanti} have pursued further developments for
these bouncing braneworlds. In particular, ref.~\cite{kanti} presents a critical
examination of the phenomenology of these cosmologies.

In the following we re-examine these bouncing brane cosmologies,
paying careful attention to the global structure of the bulk
spacetime. We find that generically these cosmologies are in fact
singular. In particular, we show that a bouncing brane must cross
the Cauchy horizon in the bulk space. However, the latter surface
is unstable when arbitrarily small excitations are
introduced in the bulk spacetime. The remainder of the paper is
organized as follows: We review the construction of the bouncing
braneworld cosmologies in section 2. Section 3 presents a
discussion of the global structure of the full five-dimensional
spacetime and the instability associated with the Cauchy horizon.
We conclude in section 4 with a brief discussion of our results.

\section{Construction of a Bouncing Braneworld}

We consider a four-dimensional brane coupled to five-dimensional
gravity with the following action
\beq {\cal{I}}=\frac{1}{16\pi G_5}\int_{\cal M}
d^5x\sqrt{-g}\left[{R}_5+ \frac{12}{L^2}-{1\over4} F^2\right] -
\frac{3}{4\pi G_5 \lambda}\int_{\cal B} d^4 x \sqrt{-\gamma}+
\int_{\cal B} d^4 x \sqrt{-\gamma}{\cal L}\, . \labell{action}\eeq
Here, ${R}_5$ denotes the Ricci scalar for the bulk metric, $g_{\mu \nu}$,
and $F_{\mu \nu}$ is the field strength of a bulk gauge field. The
(negative) bulk cosmological constant is given by
$\Lambda_5=-6/L^2$, while the brane tension is $T=\frac{3}{4\pi G_5
\lambda}$. The length scales $L$ and $\lambda$ are introduced here
to simplify the following analysis. The induced metric on the
brane is denoted by $\gamma_{ab}$. With the last term in the
action \reef{action}, we have allowed for the contribution of
extra field degrees of freedom which are confined to the brane,
\eg the Standard Model fields in a RS2 scenario \cite{RS2}.

The bulk equations of motion are satisfied by the five-dimensional
charged AdS black hole solution with metric
\beq ds_5^2=-V(r)\,dt^2+{dr^2\over V(r)}+r^2d\Sigma_{k}^2\ ,
\labell{solution}\eeq
where
\beq V(r)\equiv \frac{r^2}{L^2}+k-\frac{\mu}{r^2}+
\frac{q^2}{3r^4} \labell{vee}\eeq
and the gauge potential is $A_t = {q\over 2r^2}$.
In the metric above, $d\Sigma_{k}^2$ denotes the line element on a
three-dimensional sphere, flat space or hyperbolic plane for
$k=+1,0$ or --1, respectively (with unit curvature for the cases
$k=\pm1$). The parameters $\mu$ and $q$ appearing in the solution
are related to the ADM mass and charge of the black hole --- see,
\eg \cite{kanti,cata}. Note that this solution contains a
curvature singularity at $r=0$, but if $\mu$ is large enough,
there are two horizons at radii $r=r_\pm$ solving $V(r_\pm)=0$. A
Penrose diagram illustrating the maximal analytic extension of
such a black hole spacetime is given in Figure \ref{penrosediag1}.
In different parameter regimes, the positions of these two
horizons may coincide (or vanish, \ie $r_\pm$ become complex) to
produce an extremal black hole (or a naked singularity). We will
not consider these cases in the following.
\FIGURE{
     \resizebox{6cm}{9.75cm}{\includegraphics{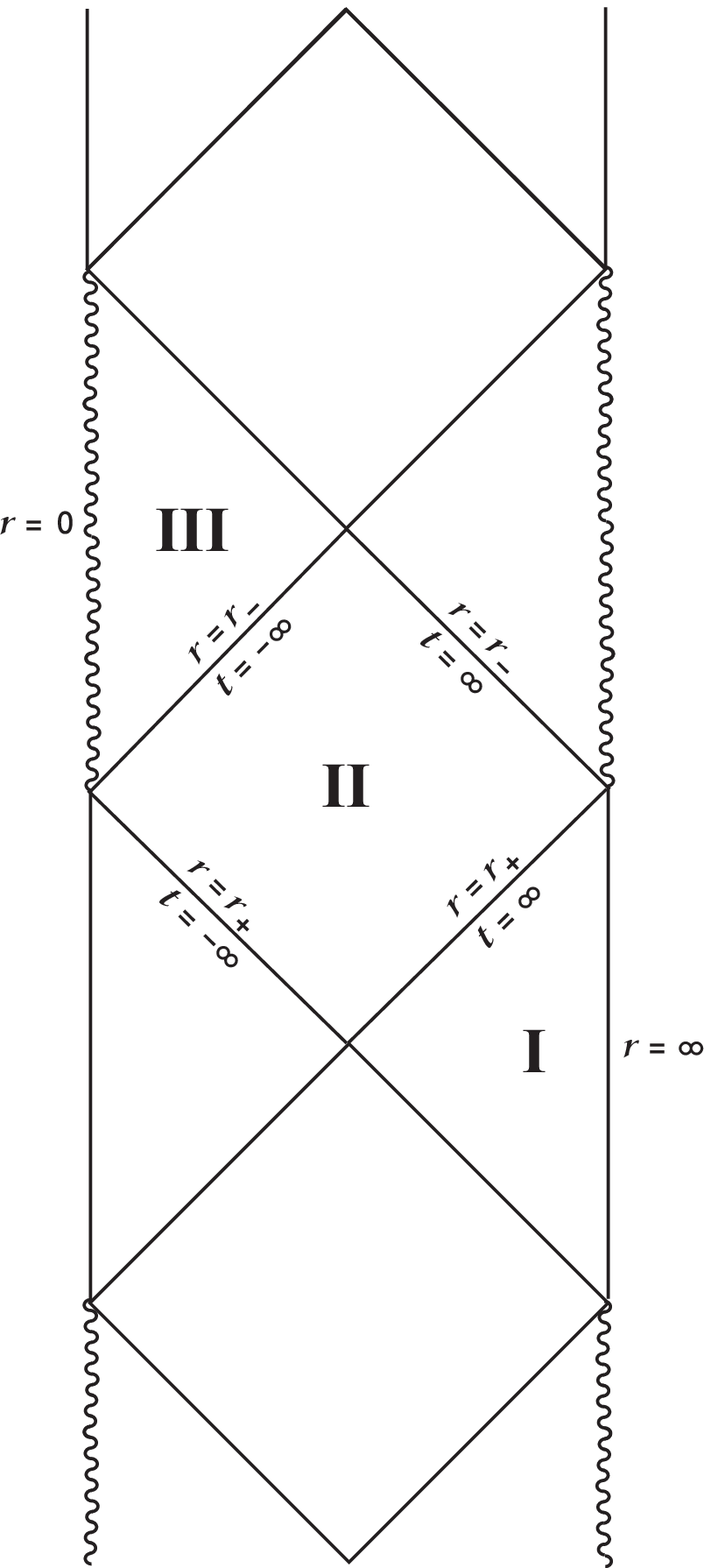}}
     \caption{Penrose diagram for maximally extended AdS
              Reissner-Nordstr\"{o}m black hole.}
     \label{penrosediag1}
}

The brane is modelled in the usual thin-brane approximation. That
is, its   worldvolume is a hypersurface, $\cal B$, which
divides the bulk spacetime, $\cal M$, into two regions. At this
hypersurface, the bulk metric is continuous but not
differentiable. Using the standard Israel junction conditions
\cite{Israel} (see also \cite{MTW}), the discontinuity in the
extrinsic curvature is interpreted as a $\delta$-function source
of stress-energy due to the brane. Then, defining the discontinuity
in the extrinsic curvature across $\cal B$ as ${\cK}_{ab}\equiv
K^+_{ab}-K^-_{ab}$, the surface stress-tensor is given by
\beq S_{ab}={1\over8\pi G_5}\left(\cK_{ab}-\g_{ab}\,
\cK^c{}_c\right)\ . \labell{surfT} \eeq
In the case of an empty brane with only tension (\ie a brane on
which no internal degrees of freedom are excited), one has
$S_{ab}=-T\g_{ab}$.

The construction of the braneworld cosmology \cite{KR} then
proceeds by taking two copies of the AdS black hole geometry,
identifying a four-dimensional hypersurface $r=a(\tau)$,
$t=b(\tau)$ in each, cutting out the spacteime regions beyond
these hypersurfaces and `gluing' the two remaining spacetimes
along these surfaces. While asymmetric constructions are possible
(see \eg \cite{kanti}), we will focus on the case where the two
bulk spacetime geometries are characterized by the same physical
parameters ($\mu,q,L$). With this choice, the calculation of the
surface stress-tensor simplifies, since ${\cal K}_{ab}=2K^+_{ab}$.
Note, however, that the gauge fields are chosen with
opposite signs on either side of the brane. Then the flux
lines of the bulk gauge field extend continuously over the brane,
starting from a positively-charged black hole on one side and
ending on the negatively-charged one on the other. In this
case, the brane carries no additional charges. We will return to
consider a charged brane in the discussion section. Since the
black hole geometry includes two separate, asymptotically AdS
regions, an economical approach to this construction would be to glue
together two mirror surfaces in each of the asymptotic
regions.\footnote{Note that this periodic construction is
distinct from the RS1 models \cite{RS1}, \eg there is a single
positive tension brane here, rather than two branes one of which has
a negative tension.}

Of course, the hypersurface described above must be determined to
consistently solve the Einstein equations (or alternatively, the
Israel junction conditions \reef{surfT}) for a physically
reasonable surface stress-tensor. Here we follow the analysis of
ref.~\cite{KR}. Identifying the time coordinate on the brane as
the proper time, $\tau$, fixes
\beq V(a)\,\dot{b}^2={\dot{a}^2\over V(a)}+1\ .\labell{bequ}\eeq
The induced metric then takes a standard FRW form:
\beq ds^2=\g_{ab}\,dx^a dx^b=-d\tau^2+a(\tau)^2d\Sigma_{k}^2\ .
\labell{indg}\eeq
Again, the brane worldvolume in the bulk spacetime \reef{solution}
is given by $r=a(\tau)$ and $t=b(\tau)$ and so the Israel
junction conditions \reef{surfT} imply
\beq
\frac{\left(V(a)+\dot{a}^2\right)^{1/2}}{a}=\frac{1}{\lambda}+\frac{4\pi
G_5}{3}\rho\ , \labell{bound} \eeq
where the `dot' denotes $\partial_\tau$, and we have included a
homogenous energy density $\rho$ for brane matter. Stress-energy
conservation would imply that the latter satisfies $\dot{\rho}+
3\frac{\dot{a}}{a}\,(\rho+p)=0$, where $p$ is the pressure due to
brane matter.

A conventional
cosmological or FRW constraint equation for the
evolution of the brane is produced by squaring eq.~\reef{bound}:
\beq \left(\frac{\dot{a}}{a}\right)^2+\frac{k}{a^2}=
{1\over\ell^2}+\frac{\mu}{a^4}-\frac{q^2}{3a^6}+
\left(\frac{1}{T\lambda}\right)^2\left(2T \rho+\rho^2\right)\ .
\label{FRW}
\end{equation}
Here, we have defined a `vacuum' curvature scale on the brane as
\beq \frac{1}{\ell^2}\equiv \frac{1}{\lambda^2}- \frac{1}{L^2}\ .
\labell{brancurv}\eeq
Implicitly, $\ell^2$ is assumed to be positive here, which leads
to the cosmological evolution being asymptotically de Sitter.
However, this assumption is inconsequential for analysis of the
cosmological bounce which follows. We can also write out the
effective cosmological and Newton constants in the
four-dimensional braneworld as
\beq \Lambda_4\equiv{3\over8\pi G_4}{1\over\ell^2}
=\frac{T}{2}\left (1-\left (\frac{\lambda}{L}\right)^2 \right)\ ,
\qquad\qquad G_4 \equiv\frac{G_5}{\lambda}\ ,\labell{DEFS} \eeq
where the latter comes from matching the term in eq.~\reef{FRW}
linear in $\rho$ to the conventional FRW equation in four
dimensions: $\left (\frac{\dot{a}}{a} \right
)^2+\frac{k}{a^2}=\frac{8\pi G_4}{3} \rho$. Of course, the FRW
constraint in this braneworld context also comes with an
unconventional term quadratic in $\rho$ \cite{CS}.

The bulk
geometry introduces various sources important in the
cosmological evolution of the brane. The mass term, $\mu/a^4$,
behaves like a conventional contribution coming from massless
radiation. The charge term, $-q^2/a^6$, introduces a more exotic
contribution with a {\it negative} energy density. This is another
example of the often-noted result that the bulk contributions to
the effective stress-energy on the brane \cite{stress} may be
negative --- see, \eg \cite{negative}.

Many exact and numerical solutions for the Friedmann equation
\reef{FRW} can be obtained in various situations, \eg
\cite{others,higherd}. However, one gains a qualitative intuition
for the solutions in general by rewriting eq.~\reef{FRW} in the
following form:
\beq 0=\dot{a}^2+U(a)\ , \labell{ham} \eeq
where
\beqa
U(a)&=&V(a)-\frac{a^2}{\lambda^2}\labell{ueff}\\
&=&k-{\mu\over a^2}+{q^2\over 3a^4} -{a^2\over\ell^2}
\nonumber\eeqa
and for simplicity we have assumed an empty brane, \ie
$\rho=0$. In this form, we recognize the evolution equation as the
Hamiltonian constraint for a classical particle with zero energy,
moving in an effective potential $U(a)$. In this case, the
transition regions where the braneworld cosmology `bounces' are
identified with the turning points of the effective potential. We
have also expressed the latter in terms of the metric function $V(a)$
in eq.~\reef{ueff} because we will want to discuss the position of
the turning points relative to the position of the `horizons', \ie
$r_\pm$. Recall that we assume the bulk solution corresponds
to a black hole with a nondegenerate event horizon. That is, we
will assume that there are two distinct solutions, $r_\pm$, to
$V(r)=0$. Then, there are two physically distinct possibilities for
a bounce.
\FIGURE{
     \resizebox{7cm}{5cm}{\includegraphics{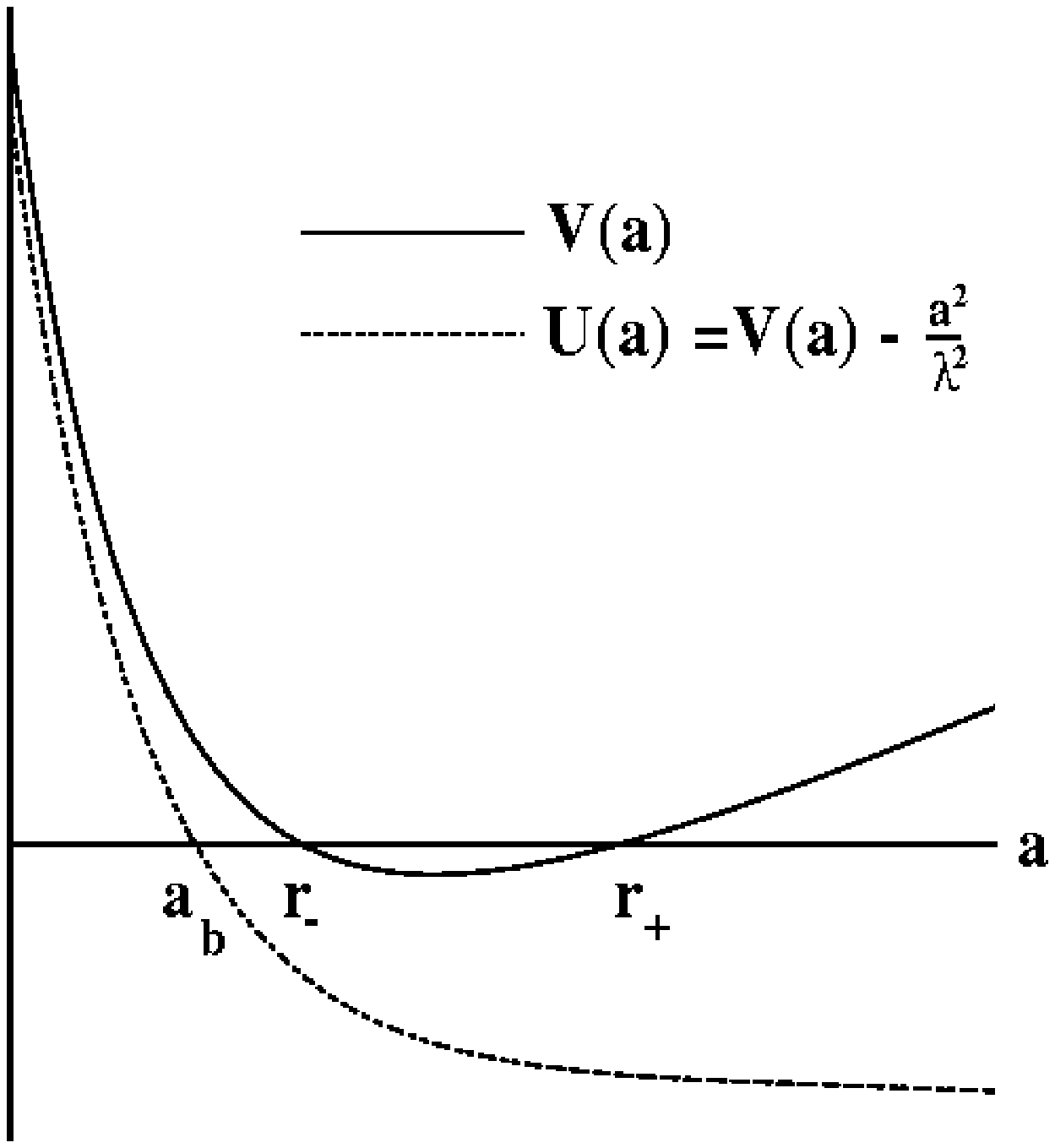}}
     \caption{Effective potential, $U(a)$, appearing in eq.~\reef{ham} for
     the evolution of the scale factor, $a(\tau)$. The turning point, $a_b$,
     occurs inside the Cauchy horizon $r_-$.}
     \label{potentials}
}

The first only occurs with $k=+1$, \ie with a spherical brane
world, and positive $\ell^2$ (or equivalently $\L_4>0$). In this
case, at large $a$, the effective potential becomes large and
negative. The next most important contribution at large $a$ is the
constant term $k$ and hence if $k=+1$, the potential may have a
zero at large $a$. This bounce is typical of those one
might find in a de Sitter-like spacetime, \eg \cite{tall}. It is
driven by the spatial curvature and occurs as long as the
effective energy density from the bulk black hole contributions or
braneworld degrees of freedom is not too large. The turning point
occurs at some large $a_{dS}$ and in particular, it is not
difficult to show that $a_{dS}>r_+$. That is, the brane bounces
before reaching the black hole. In fact the presence of the black
hole with or without charge is really irrelevant to this kind of
bounce. For example, setting $\mu=0=q$ in eq.~\reef{ueff}
produces a de Sitter cosmology on the empty brane.

The second type of bounce is generic for a wide range of
parameters. It occurs at small $a$ where the positive $q^2/a^4$
term dominates the potential \reef{ueff}, \ie where the exotic
negative energy dominates the Friedmann constraint \reef{FRW}. As
is clear from the first line of eq.~\reef{ueff}, $U(a)<V(a)$ and
therefore the turning point occurs at $a_b$, inside the position of
the Cauchy horizon, \ie $a_b<r_-$, as illustrated in Figure
\ref{potentials}. The latter result will be essential in the
following discussion.

The Penrose diagram for the bouncing braneworld cosmologies is
shown in Figure \ref{penrosediag2}. In the `cut and paste'
procedure outlined above, the singularity on the right side of the
first black hole is cut out, but the singularity on the left
remains. Hence, the remaining portion of the $r=r_-$ surface is still a
Cauchy horizon.
\FIGURE{
     \resizebox{10cm}{8cm}{\includegraphics{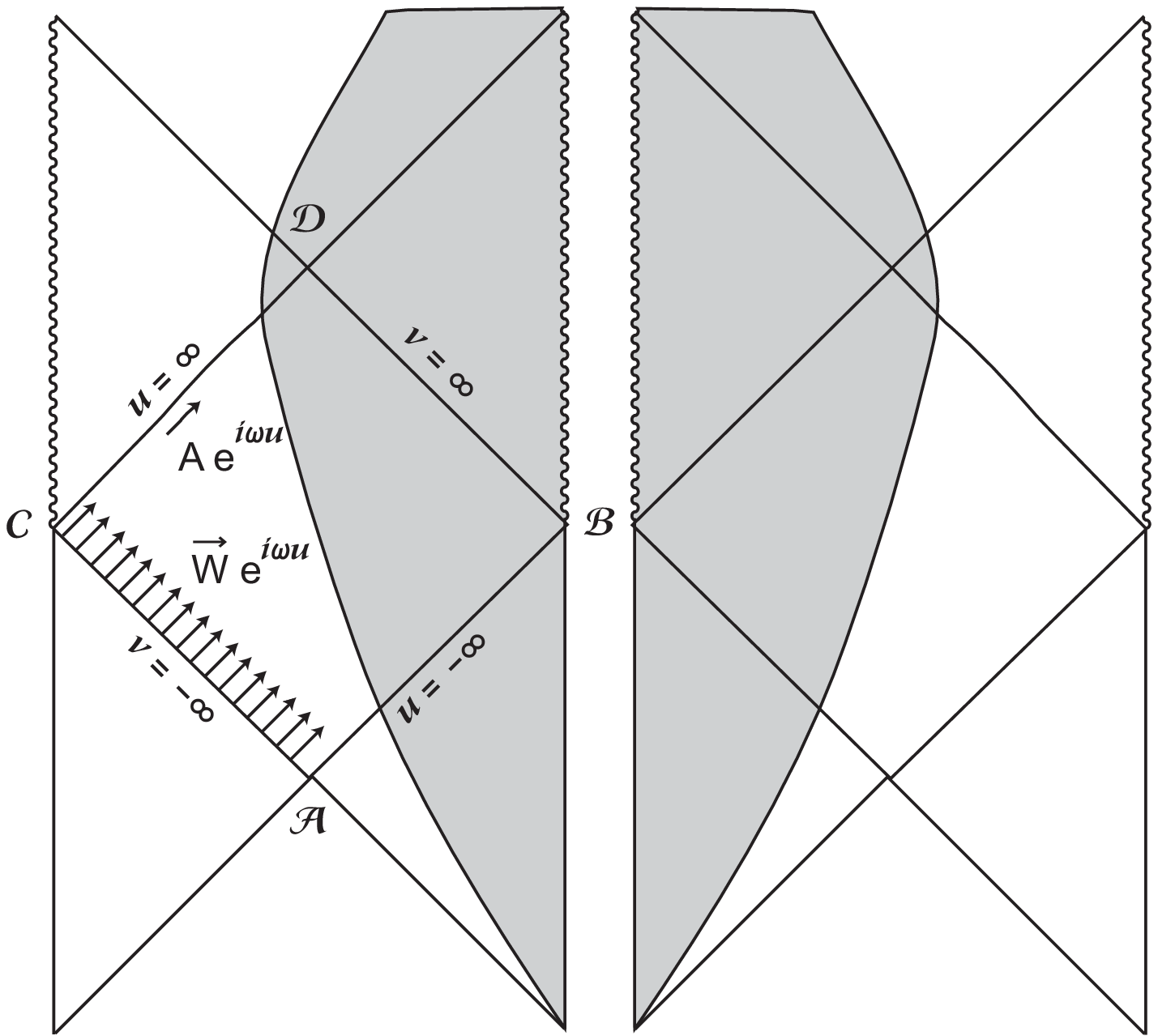}}
     \caption{Penrose diagram for a bouncing braneworld model with ingoing modes
at the event and Cauchy horizons.  The grey areas are those regions of
spacetime that are cut out in the construction, with identification
performed along the boundary.}
     \label{penrosediag2}
}

Note that in Figure \ref{penrosediag2}, the brane trajectory
enters the region between the horizons across the segment
$\mathcal{AB}$ and exits across the opposite segment
$\mathcal{CD}$. One can verify that this occurs in all cases using
eqs.~\reef{bound} and \reef{bequ}. From the latter, we find that
\beq \dot{t}=\pm\frac{a}{V(a)}\left (\frac{1}{\lambda}+\frac{4\pi
G_5}{3}\rho \right )\ . \labell{righto}\eeq
If the brane tension and the energy density, $\rho$, are both
assumed to be always positive, then the last factor is always
positive. Furthermore, for $r_- < a < r_+$, $V(a)<0$. Hence, the
right hand side above is non-zero and has a definite sign for the
entire range $r_- < a < r_+$. Therefore $\dot{t}$ cannot change
sign along the brane trajectory within the black hole interior. It
then follows that if a trajectory starts at a point on
$\mathcal{AB}$ with $(t,r)=(\infty,r_+)$, then it must run across
the black hole interior to a point on $\mathcal{CD}$ with
$(t,r)=(-\infty,r_-)$ --- see Figures \ref{penrosediag1} and
\ref{penrosediag2}.

\section{Instability Analysis}

In the previous section, we reviewed the construction of a broad
family of bouncing braneworld cosmologies \cite{PEL}. A key result
was that the turning point for the brane's trajectory in the bulk
geometry was inside the Cauchy horizon of the charged AdS black
hole. However, previous studies in classical general relativity
found that the Cauchy horizon is
unstable when generic perturbations are introduced for charged
black holes in asymptotically flat \cite{chandrasekhar}, or de
Sitter \cite{dsrn1,dsrn2} spaces. Below, we will show that the same
instability arises in the asymptotically AdS case as well.
This is problematic for the bouncing braneworld cosmologies, as
generically the contracting brane will reach a curvature
singularity before it begins re-expanding.

In the following, we demonstrate the instability of the Cauchy
horizon to linearized perturbations in the bulk. Our approach will
be two-fold. We begin by examining linearized fluctuations of a
massive Klein-Gordon field propagating in the background.
Secondly, we consider gravitational and electromagnetic
perturbations. In both cases, it is found that an observer
crossing the Cauchy horizon would measure an infinite flux from
these modes. The expectation is then that the full nonlinear
evolution, including the back-reaction on the background metric,
will produce a curvature singularity.

Many of the expressions appearing in the linearized analysis
involve the surface gravities of the two horizons in the
background. The surface gravities are given, as usual, by
\beq \kappa_\pm = \frac{1}{2} \left | \frac{d V}{d r} \right
|_{r=r_\pm}\ . \labell{surfgravities} \eeq
An important observation in the following is that
$\kappa_+/\kappa_-<1$, which follows from $r_-<r_+$. Now it will
be convenient to define the event and Cauchy horizons implicitly
by re-expressing the metric function \reef{vee} as
\beq V(r)=\frac{(r^2-r_+^2)(r^2-r_-^2)(r^2+r_0^2)}{L^2r^4}\ .
\labell{newvee} \eeq
This expression also defines $r_0$ as determining the complex
roots of $V(r)$. Further, the analysis is facilitated by
introducing some new coordinates to describe the background
geometry \reef{solution}. In particular, we define the radial
tortoise coordinate
\beq r_* \equiv \frac{1}{2\kappa_+}\log \frac{|r-r_+|}{r+r_+}
-\frac{1}{2\kappa_-}\log \frac{|r-r_-|}{r+r_-}
+\frac{r_0^3L^2}{(r_0^2+r_-^2)(r_0^2+r_+^2)} \tanh^{-1}
\frac{r}{r_0}\ , \labell{tort} \eeq
which is chosen to satisfy $dr_*=dr/V(r)$. The focus of the
following analysis will be the behavior of linearized
perturbations in the range $r_-<r<r_+$ (\ie region II in Figure
\ref{penrosediag1}). In this region, we have
$r_*\rightarrow\pm\infty$ as $r\rightarrow r_\mp$. Finally, it will
be useful to work with null coordinates,
\beq u=r_*-t\ ,\qquad v=r_*+t\ ,\labell{null}\eeq
with which the line element becomes
$ds^2=V(r)du\,dv+r^2d\Sigma^2_k$.

The massive Klein-Gordon equation in the charged black hole
background \reef{solution} may be expanded as
\beq -\frac{1}{V(r)} \partial_t^2 \Phi
+V(r) \partial_r^2 \Phi
+\frac{1}{\sqrt{-g}}\partial_r \left( \sqrt{-g} ~V(r) \right ) \partial_r \Phi
+\frac{1}{r^2}\nabla_k^2 \Phi-M^2 \Phi=0 \ , \eeq
where we write $\nabla^2_k$ for the Laplacian on the
three-dimensional space $\Sigma_k$ appearing in the line element
\reef{solution}. The eigenvalue problems for $\partial_t^2$ and
$\nabla_k^2$ each have known solutions with eigenvalues, say,
$-\omega^2$ and $-n^2_k$. Hence by separation of variables, the
Klein-Gordon equation is reduced to a single ordinary differential
equation for $\Phi(r)$,
\beq - \partial_{r_*}^2\tilde{\Phi} +V(r)\left
(\frac{n_k^2}{r^2}+M^2+\frac{3V'}{2r}+\frac{3V}{4r^2} \right )
\tilde{\Phi}=\omega^2\,\tilde{\Phi} \ , \label{kleingordon} \eeq
where we have introduced the tortoise coordinate \reef{tort} and
rescaled $\tilde{\Phi}=r^{3/2} \Phi$.

As we approach the Cauchy horizon, the second term on the left
hand side of eq.~\reef{kleingordon} vanishes, leading to
oscillatory solutions $\exp(\pm i\omega r_*)$. Now, the flux seen
by an observer freely falling across the horizon, with
five-velocity $U^\mu$, is proportional to the square of the scalar
${\cal F}=U^{\mu}\partial_\mu\Phi$. $\cal F$ then includes a
contribution proportional to $e^{\kappa_-u
}\partial_u\tilde{\Phi}$ near the Cauchy horizon. Since the
solutions of eq.~\reef{kleingordon} are oscillatory as
$r\rightarrow r_-$, we have that this term, and hence ${\cal F}$,
diverges. Similar divergences appear in the observed energy
density for these linearized perturbations, and so the expectation
is that when back-reaction is included, the metric will develop a
curvature singularity.

Next we proceed to a more rigorous analysis of metric and Maxwell
field perturbations, following the method of Chandrasekhar and
Hartle \cite{chandrasekhar}. We are simply establishing the
existence of unstable modes and so, for simplicity, we fix $k=0$
and consider an ``axial'' perturbation of one of the flat space
coordinates. However, the extension of this analysis to general
perturbations and backgrounds is straightforward.

The unperturbed bulk metric \reef{solution} is
\beq
ds^2=-V(r)\,dt^2+\frac{dr^2}{V(r)}+\frac{r^2}{L^2}(dx^2+dy^2+dz^2)
\ , \labell{soluteb}\eeq
where $V(r)$ is as given in eq.~\reef{vee} with $k=0$. We now
focus on a class of perturbations where this metric is modified by
replacing
\beq
dz \rightarrow dz+q_t(t,r) dt+q_r(t,r) dr\ .
\labell{axial}
\eeq
Similarly
for the Maxwell field, we introduce perturbations:
$\delta {\mathbf F}=(f_{tz}(t,r) {\mathbf d}t+f_{rz}(t,r) {\mathbf
d}r) \wedge {\mathbf d}z$. The linearization of the bulk Einstein
and Maxwell equations about the background solution may be reduced
to a single Schr\"{o}dinger-like equation:
\beq -\partial_{r_*}^2 F + \frac{V}{r}\left(\frac{4q^2}{r^5}
-\frac{V}{4r}+\frac{V'}{2} \right) F= \omega^2\,F \ .
\labell{schrodinger} \eeq
In this equation, we have defined $F\equiv r^{1/2} f_{tz}$ and
assumed an $e^{-i\omega t}$ dependence for all fields. To apply
the standard results of scattering theory below, it is important
to note that the effective potential (\ie the pre-factor in the
second term on the left hand side) is bounded, negative and
integrable throughout the black hole interior. Further, we note
that the effective potential vanishes as $ \exp \left ({\pm
\kappa_{\pm} r_*} \right)$ for $r_* \rightarrow \mp \infty$. The
other components of the perturbation are related to $F$ by
\bea
f_{rz} & = & \frac{i}{\omega}\partial_r (F/r^{1/2})\ , \nonumber \\
Q_{tr} & = & -4i\frac{q\,L^2}{\omega\,r^{11/2}}F\ ,
\labell{perts}\eea
where $Q_{tr}=\partial_t
q_{r}-\partial_r q_t$. Note that the linearized equations only fix
the metric perturbations, $q_{r}$ and $q_t$, up to infinitesimal coordinate
transformations of $z$, but $Q_{tr}$ provides a gauge invariant
combination which is completely determined.

We introduce a basis of solutions to eq.~\reef{schrodinger} normalized
so that near the event horizon, \ie $r_*
\rightarrow -\infty$,
\beqa
\stackrel{\rightarrow}{F}(\omega,r_*) & \rightarrow & e^{i \omega r_*}\ , \nonumber \\
\stackrel{\leftarrow}{F}(\omega,r_*) & \rightarrow & e^{-i \omega
r_*}\ , \label{functions1} \eeqa
representing, respectively, ingoing and outgoing modes. The full
solution to \reef{schrodinger} may then be written as
\beq F(t,r_*)=\int^\infty_{-\infty}
\frac{d\omega}{2\pi} \left [ \stackrel{\leftarrow}{W}(\omega)
\stackrel{\leftarrow}{F}(\omega,r_*)
+\stackrel{\rightarrow}{W}(\omega)
\stackrel{\rightarrow}{F}(\omega,r_*) \right ] e^{-i\omega t}\ .
\label{fullsoln}
\eeq
At present, we are only interested in the ingoing modes, whose
profile is determined by $\stackrel{\rightarrow}{W}(\omega)$. The
outgoing modes may be similarly dealt with, but extra analysis
would be required to show they lead to a divergent flux.
We will return to this point near the end of the section.

We are free to choose any reasonable initial profile for the ingoing modes.
However, one restriction which we impose on the initial frequency distribution
$\stackrel{\rightarrow}{W}(\omega)$ of ingoing modes is
that an observer falling across the event horizon at $\mathcal{AB}$
measures a finite flux.  The flux for such an observer
contains a term $\mathcal{F} \sim e^{-\kappa_+ u}\partial_u F$. Hence
considering eq.~\reef{fullsoln}, we require that $\stackrel{\rightarrow}
{W}(\omega)$ have at least
one pole with $\mathfrak{Im}(\omega) \le -\kappa_+$.

The initially-ingoing modes are scattered by the potential in region II,
leading to both ingoing and outgoing modes at the Cauchy horizon.  Scattering theory
will be used to relate the functions introduced in \reef{functions1} to a second
set normalized so that, as $r_* \rightarrow \infty$,
\beqa \stackrel{\rightarrow}{F}(\omega,r_*) e^{-i\omega t}
 & \rightarrow &
\left(A(\omega) e^{i\omega r_*}
          +B(\omega) e^{-i \omega r_*}\right) e^{-i\omega t}\  \nonumber \\
&  &\quad= A(\omega)e^{i \omega u}+B(\omega)e^{-i \omega v}\ .
\labell{skatter}
\eeqa
Clearly, the dominant contribution to the flux at the Cauchy horizon
results from
\beq
{\mathcal F} \sim e^{\kappa_- u}
\int^{\infty}_{-\infty} d\omega \ \omega \stackrel{\rightarrow}{W}(\omega)
A(\omega)e^{i \omega u}\ .
\label{fluxintegral}
\eeq
In terms of the Schr\"{o}dinger equation describing the
perturbations, it is the modes that are ``transmitted'' across the
potential that constitute this potentially-divergent flux.  These
modes skim along just outside the Cauchy horizon heading towards
the brane.

The integral in \reef{fluxintegral} may be computed by closing the
contour in the upper-half-plane.  Using arguments from
\cite{chandrasekhar}, with virtually no modification, we find that
$\omega A(\omega)$ is analytic in the infinite strip between $\pm
i \kappa_+$.  For simplicity, we'll further assume that
$\stackrel{\rightarrow}{W}(\omega)$ is analytic in the strip
$[0,i\kappa_+]$ and that it is non-zero for $\omega=i\kappa_+$.
With these assumptions, the leading term in \reef{fluxintegral} is
from the residue of the pole at $\omega=i \kappa_+$:
\beq {\mathcal F} \sim e^{(\kappa_- -
\kappa_+)u} \left \{i \kappa_+ \stackrel{\rightarrow}{W}(i
\kappa_+) 2\pi i~ \mbox{Res} \left (A(i\kappa_+)\right) \right \}\
.
\label{finalfluxintegral}
\eeq
Since $\kappa_- >\kappa_+$, this flux always diverges as $u
\rightarrow \infty$. Relaxing our assumptions on the analyticity
of $\stackrel{\rightarrow}{W}(\omega)$ in the upper half plane
could lead to additional divergent contributions to the flux, but
we will not consider those here.

Note that the brane and boundary conditions at the brane played no
role in the scattering analysis above. While the brane will affect
the complete scattering of modes inside the event horizon, the
basic source of the instability is the same piling up of infalling
modes on the Cauchy horizon found in previous examples
\cite{chandrasekhar,dsrn1}. Hence we disregard the details of the
scattering of modes at the brane, just as the original discussion
of the instability for the Reissner-Nordstr\"{o}m black hole
\cite{chandrasekhar} ignored the presence of a collapsing star
forming the black hole.

However, for completeness, let us briefly discuss the boundary
conditions which must be imposed on the perturbations at the
brane. First, the metric perturbations must be matched across the
brane surface so that no additional contributions are induced in
the surface stress-energy \reef{surfT}. In particular, the axial
perturbation \reef{axial} considered above induces a new $K_{\tau
z}$ component in the extrinsic curvature, and this component must
be continuous across the brane. Similarly, continuity is imposed
on the Maxwell field strength. More precisely, to ensure that no
electric charges or currents are implicitly induced on the brane,
we require that all components $n^\mu F_{\mu\nu} t^\nu$ are
continuous, where $n^\mu$ and $t^\nu$ are the unit normal and any
tangent vector to the brane.  Finally, since we are working with
perturbations to the field strength directly, and not the gauge
potential, we must demand continuity of the tangential components,
$t_1^\mu F_{\mu \nu} t_2^{\nu}$, to ensure there are no magnetic
charges or currents induced, \ie $\mathbf{dF}=0$.

We close this section with a discussion of the initially-outgoing
modes defined by the distribution
$\stackrel{\leftarrow}{W}(\omega)$ in eq.~\reef{fullsoln}. In
Figure \ref{penrosediag2}, we will primarily consider modes
entering the interior region on the left through the lower portion
of $\mathcal{AB}$. In this case, to contribute to the instability
at the Cauchy horizon $\mathcal{CD}$, these modes must be
reflected by the curvature (\ie by the effective potential in
eq.~\reef{schrodinger} to become ingoing. This scattering leads to
a different analytic structure in eq.~\reef{skatter} for
$A(\omega)$, describing the $u$-dependent modes at the Cauchy
horizon. In the ``un-cut'' spacetime with no brane in place, this
structure is identical to that obtained for the contribution of
the initially-ingoing modes to $B(\omega)$. Of course, inserting
the brane in the black hole interior produces a more complicated
scattering problem, the details of which would depend on the
precise brane trajectory. For example, the outgoing flux would
receive additional contributions from perturbations transmitted
across the brane from the right hand side of Figure
\ref{penrosediag2}, as well as from initially-ingoing modes which
are back-scattered by the brane. We did not attempt a detailed
study of these contributions.

Now , following the standard analysis with no brane in place, we
find that the contribution to $\omega A(\omega)$ from the outgoing
modes is analytic in the semi-infinite strip
$(-i\kappa_+,i\kappa_-)$.  {If} we assume that
$\stackrel{\leftarrow}{W}(\omega)$ is analytic in the strip
$(0,i\kappa_-)$, {then} we would find, upon closing the contour in
the upper-half-plane, that the contribution to the flux is finite.
However, it is consistent with the requirement that an observer
crossing $\mathcal{AC}$ measure a finite flux, to allow
$\stackrel{\leftarrow}{W}(\omega)$ to have poles in the range
$\kappa_+ \le \mathfrak{Im}(\omega)<\kappa_-$. With such a choice,
there will be divergent contributions to the flux, provided that
the residue of $A$ is non-zero at these poles. This effect differs
from that discussed above in that the leading contribution to the
flux comes from a pole in the initial frequency distribution
rather than the scattering coefficient $A(\omega)$. A similar
discussion played an important role in demonstrating the
instability of the Cauchy horizon of de
Sitter-Reissner-Nordstr\"{o}m black holes over the entire range of
physical parameters \cite{dsrn2}.

 \section{Discussion}

One of the most interesting features of the braneworld cosmologies
presented in ref.~\cite{PEL} is that, while they seem to evade any
cosmological singularities, their evolution is still determined by
a simple effective action, albeit in five dimensions. However, our
present analysis indicates that instabilities arise in the
five-dimensional spacetime, and that the brane will generically
encounter a curvature singularity before bouncing. The two
essential observations leading to this result were: i) the turning
point for the brane cosmology occurs inside the Cauchy horizon of
the maximally-extended geometry of the charged AdS black hole and
ii) a standard analysis within classical general relativity shows
that the Cauchy horizon is unstable against even small excitations
of the bulk fields. Note that from these results we cannot
conclude that the brane does not bounce, but rather due to the
appearance of curvature singularities, the evolution can not be
reliably studied with the original low energy action
\reef{action}.

Of course, one may ultimately have reached this conclusion since
the full bulk spacetime still includes a curvature singularity at
$r=0$ --- see Figure \ref{penrosediag2}. However, while the latter
remains distant from the brane,  those at the Cauchy horizon are
of more immediate concern as they intersect the brane's
trajectory.

In the discussion of metric and gauge field perturbations in
section 3, we fixed $k=0$ and limited ourselves to modes that
depended only on $t$ and $r$ to simplify the discussion. One may
be concerned by the fact that these modes have infinite extent in
the three-dimensional flat space and so we present a brief
discussion of the full analysis.  Generalizing our results to the
most general perturbation is straightforward but tedious. For an
arbitrary linearized perturbation, the separation of variables
would naturally lead to considering Fourier components in the
($x,y,z$) directions with a factor $\exp \left(i \vec{n} \cdot
\vec{x} \right )$.  Since we require a superposition of these
modes for many different $\vec{n}$ to localize the perturbation,
we cannot simply rotate in the flat space to remove the
dependence on one of the spatial coordinates.  Thus the general
analysis necessarily involves an ansatz for the perturbations
dependent on all five coordinates, which, of course, requires
extending the perturbations to additional components of both the
metric and gauge field. Appropriate linear combinations of these
perturbations would decouple, giving a set of Schr\"{o}dinger-like
equations, similar to that found above. While the potentials in
each of these equations is different, there are typically simple
relationships between them implying relations between the
solutions --- for further discussion of these relations, see
\cite{chandrasekhar}. Then it is sufficient to solve only one of
the equations, and the analysis, and the results, are essentially
the same as presented above

Of course, our preliminary analysis with massive Klein-Gordon
modes included all of the spatial modes, and further applied for
all of the possible values of $k$, specifying the spatial
curvature on the brane. In all cases, there was an infinite flux
of these modes at the Cauchy horizon. While further analysis of
the full scattering and boundary conditions would be required to
make this consideration of fluxes rigorous, the end result would
be the same. Hence we are confident that the results for the
metric and gauge field perturbations with $k=0$ also carry over
for $k=\pm 1$.

Recall that, as discussed in section 2, an apparently economical
approach to constructing these bouncing cosmologies would be to
cut and paste along two mirror surfaces in each of the separate
asymptotically AdS regions of the black hole geometry. In such a
periodic construction the nature of the singular behavior would be
slightly different. As discussed around eq.~\reef{righto}, the
brane trajectory is unidirectional in the coordinate time, $t$.
Hence in Figure \ref{penrosediag2}, if a brane enters the event
horizon to the right of the bifurcation surface ${\mathcal A}$,
then it must exit through the Cauchy horizon to the left of
${\mathcal D}$. However, the same result requires that a brane
trajectory entering to the left of ${\mathcal A}$ exits to the
right of ${\mathcal D}$. Therefore in the periodic construction
above, the two mirror trajectories must cross at some point in the
region $r_-<r<r_+$, as illustrated in Figure \ref{penrosediag3}.
Hence, the evolution is singular in that the fifth dimension
collapses to zero size in a finite proper time. One redeeming
feature of this collapse is that the curvature remains finite, and
hence one might imagine that there is a simple continuation of the
evolution in which this `big crunch' is matched onto a `big bang'
geometry. Similar collapsing geometries have been a subject of
great interest in the string theory community recently --- see,
\eg \cite{slap,slap2}. Resolving precisely how the spacetime
evolves beyond such a `big crunch' is an extremely difficult
question and as yet string theory seems to have produced no clear
answer. In particular, it seems that these geometries are also
subject to gravitational instabilities \cite{wonk} not dissimilar
to those found here. In the present context, the situation is
further complicated as the precise matching procedure for the
background geometry is obscure. Naively, one might be tempted to
continue beyond the collapse point $\mathcal{E}$ with the doubly
shaded region in Figure \ref{penrosediag3}. However, a closer
examination shows that the brane would have a negative tension in
this geometry. The other natural alternative is to match the
crunch at $\mathcal{E}$ to the big bang emerging from
$\mathcal{F}$, but the gap in the embedding geometry would seem to
complicate any attempts to make this continuation precise. In any
event, it is clear that once again these knotty questions can not
be resolved using the low energy action \reef{action} alone but,
rather, one would have to embed this scenario in some larger
framework, \eg string theory. \FIGURE{
     \resizebox{9cm}{10cm}{\includegraphics{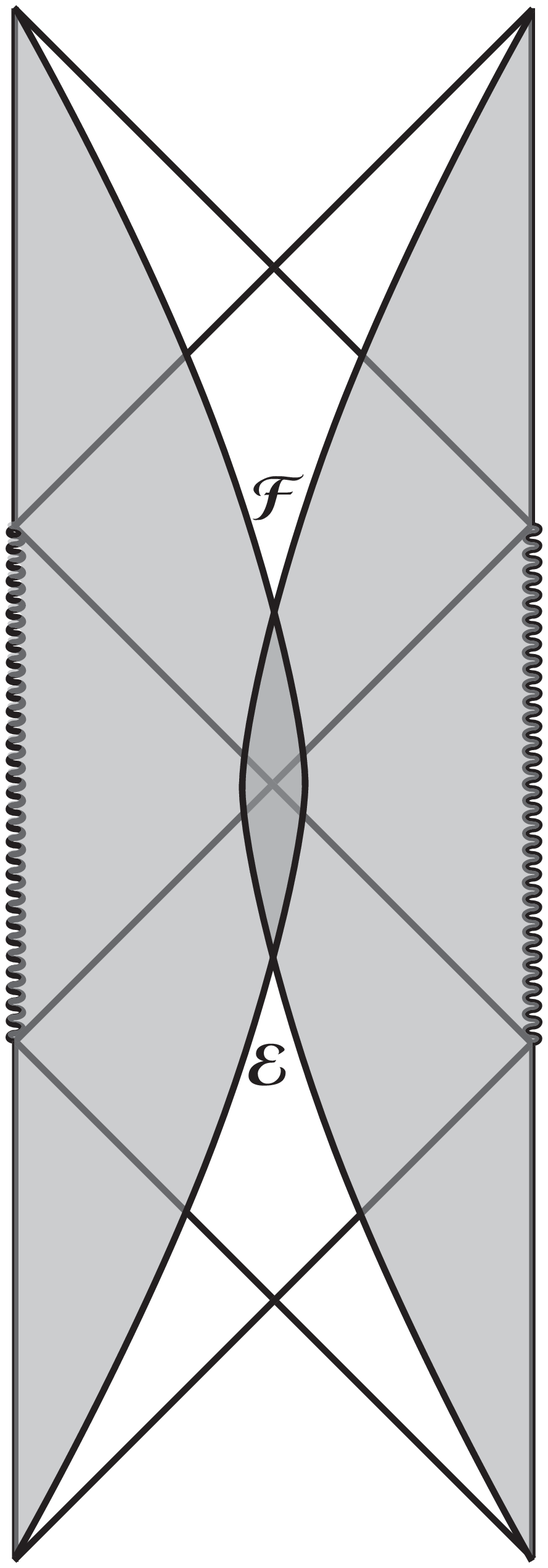}}
     \caption{Penrose diagram for periodic construction of the braneworld cosmology}
     \label{penrosediag3}
}

Much of our discussion has focussed on the bouncing cosmologies for
an empty four-dimensional brane, but the analysis and the
results are easily extended to other cases. One simple generalization
would be for higher dimensional cosmologies following, \eg
\cite{higherd}. The instability found here would also appear in the
assymetric constructions discussed in ref.~\cite{kanti}.

A more interesting generalization to consider is adding matter
excitations on the brane. As long as the energy density is
positive, such matter contributions will not affect the result
that the brane crosses the Cauchy horizon. At first sight, it
would also seem that reasonable brane matter cannot prevent the
bounce. The negative energy contribution arising from the bulk
charge is proportional to $1/a^6$. For a perfect fluid (in four
dimensions), this would require the stiffest equation of state
consistent with causality \cite{stiff}, \ie $p=\rho$. For example,
a coherently rolling massless scalar field would yield
$\rho\propto 1/a^6$. Hence it would seem that the term $-q^2/a^6$
would dominate the $\rho$ contribution coming from brane matter
and a bounce would be inevitable. However, ref.~\cite{kanti}
recently pointed out the $\rho^2$ contribution in the FRW
constraint \reef{FRW} can prevent a bounce. In fact, with any
equation of state $p=w\rho$ with $w\ge0$, this contribution can
dominate the bulk charge contribution. Hence with a sufficiently
large initial energy density on the brane, a big crunch results on
the brane. This crunch corresponds to the brane trajectory falling
into the bulk singularity at $r=0$. It then follows that the brane
must cross the Cauchy horizon in this case as well, and we expect
singularities to develop there with generic initial data.

More broadly, mirage cosmologies \cite{elias} are induced by the
motion of a brane in a higher dimensional spacetime. A general
warning which the present analysis holds for these models is that
Cauchy horizons are quite generally unstable. Hence if a
particular solution involves a brane traversing such a surface in
the bulk spacetime, one should expect that these cosmologies will
encounter singularities for generic initial data.

At this point, we observe that in the literature much of the discussion
of these brane cosmologies treats the brane as a fixed point of a $Z_2$
orbifold, rather than making a symmetric construction as discussed
in section 2. As discussed there, one must flip the sign of the gauge
potential in the background solution on either side of the brane in order
that the brane is transparent to field lines. In contrast for a $Z_2$ orbifold,
the field lines end on the brane. As there is no natural coupling of a
one-form potential to a three-brane in five dimensions, the model must be
extended to include charged matter fields on the brane. One comment
is that as the action \reef{action} does not explicitly include these
degrees of freedom or their coupling to the Maxwell field, we cannot
be sure that the analogous construction to that presented in section
2 will yield a consistent solution of all of the degrees of freedom.
One might also worry that the simplest solutions would have additional
instabilities associated with having a homogeneous charge
distribution throughout the brane.

Bouncing cosmologies have long been of interest \cite{begin}. Much
of their appeal lies in their potential to provide a calculable
framework to describe the origins of the universe. Apart from the
present discussion, braneworlds and higher dimensions have
inspired many attempts to model a bouncing cosmology, including:
pre-big bang cosmology \cite{ppb}; cyclic universes \cite{end}
based on a Lorentzian orbifold model \cite{slap};  braneworld
cosmologies induced by cyclic motion in more than one extra
dimensions \cite{Brax};\footnote{Ref.~\cite{Kachru} gives a
closely related construction embedded in string theory. Note,
however, that from the point of view of the Einstein frame in four
dimensions, there are no sources of negative energy density and
the universe is static.} universes with higher form fluxes
\cite{old,Bunce}, which are related to S-brane solutions
\cite{sbran}; braneworld cosmologies \cite{Shtanov} with an extra
internal time directions \cite{gia}. However, as well as the
present model, none of these works has yet provided a compelling
scheme which is free of pathologies or obstructions to prediction.
We may take solace from the absence of any simple bounce models in
that it appears that understanding the early universe and, in
particular, the big bang singularity demands that we greatly
expand our understanding of quantum gravity and string theory.

\acknowledgments

RCM would like to thank Renata Kallosh for bringing this problem
to his attention, and Cliff Burgess and Shamit Kachru for
interesting conversations.  For their hospitality while this paper
was being finished, RCM also thanks the KITP. Finally, we thank
David Winters for proofreading an early version of this paper.
This research was supported in part by NSERC of Canada and Fonds FCAR du Qu\'ebec.
Research at the KITP was supported in part by the National Science
Foundation under Grant No.~PHY99-07949.




\end{document}